\documentclass[aps,pra,twocolumn]{revtex4}

\usepackage{graphicx}
\usepackage{graphics}

\begin{document}

\title{Trapping fermionic $^{40}$K and bosonic $^{87}$Rb on a chip}

\author{S.\ Aubin$^1$}
\author{M.\ H.\ T.\ Extavour$^1$}
\author{S.\ Myrskog$^1$}
\author{L.\ J.\ LeBlanc$^1$}
\author{J.~Est\`{e}ve$^2$}
\author{S.\ Singh$^1$}
\author{P.\ Scrutton$^1$}
\author{D.\ McKay$^1$}
\author{R.\ McKenzie$^1$}
\author{I.\ D.\ Leroux$^1$}
\author{A.\ Stummer$^1$}
\author{J.\ H.\ Thywissen$^1$}
\homepage{http://www.physics.utoronto.ca/~jhtgroup/}
\affiliation{$^1$McLennan Physical Laboratories, University of Toronto,
Toronto, Ontario, M5S 1A7 Canada\\
$^2$Laboratoire de Photonique et de Nanostructures, UPR 20 du CNRS,
91460 Marcoussis, France}

\date{submission February 8, 2005; revision August 5, 2005}


\begin{abstract}

We demonstrate the loading of a Bose-Fermi mixture into a 
microfabricated magnetic trap.
In a single-chamber vacuum system, laser-cooled atoms
are transported to the surface of a substrate on which gold wires
have been microfabricated. The magnetic field minimum formed near these current-carrying wires is used to
confine up to $6\times10^4$ neutral $^{40}$K atoms.
In addition, we can simultaneously load $2 \times 10^5$ $^{87}$Rb atoms,
demonstrating the confinement of two distinct elements with such a trap.
In a sequence optimized for $^{87}$Rb alone, we observe up to 
$1 \times 10^7$ trapped atoms.
We describe in detail the experimental apparatus, and 
discuss prospects for evaporative cooling towards
quantum degeneracy in both species.

\end{abstract}
\maketitle


\section{Introduction}
Since the observation of Bose-Einstein condensation  \cite{BEC} and
Fermi degenerate gases  \cite{FDG}, atomic gases have been used to study
quantum degenerate many-body systems. Experiments have demonstrated
exquisite control over the trapping environment, temperature, density,
and interaction strength of the constituents. In contrast to solid state
and liquid systems, quantum degenerate gases are dilute and without impurities,
making them more easily compared to theoretical models. Furthermore, observations can
be made with spatial imaging or with momentum spectroscopy, both
of which can be done rapidly enough to observe the dynamics of the system.
These systems can be tailored to access interesting many-body phenomena
such as the Hubbard model  \cite{LukinHTSC}, Anderson
localization  \cite{Anderson}, and the BEC-BCS crossover
regime  \cite{BEC-BCS}.

Microfabrication of magnetic trapping elements brings new tools to the
study of ultra-cold quantum gases  \cite{Hansel,Ott}.  Micro-electromagnets ($\mu$EMs) can be used to build smaller traps, achieve stronger trap gradients and
curvatures  \cite{Weinstein}, and reduce the power dissipation of the
trapping elements. In addition, microfabrication techniques make integration with
waveguides  \cite{atomwg} and other devices  \cite{ReichelMotor,Folman}
possible. The ability to trap atoms near a surface has led to a range
of recent studies focusing on interactions between degenerate atom
clouds and surfaces  \cite{surfaces,Lin}. To date, $\mu$EMs have only been
used to study thermal and degenerate Bose gases. 

In this paper we discuss an apparatus built to study ultra-cold Fermi gases in a
$\mu$EM trap. The high oscillation frequency accessible in a strongly confining
$\mu$EM trap has several advantages for studies of quantum gases. First, an
increased collision rate allows fast evaporative cooling, which
simplifies the realization of a dual species system, as described in
\S\ref{sec:apparatus}. Second, the high oscillation frequency allows quantum degeneracy
with a smaller number of atoms. The Fermi
temperature in a three-dimensional harmonically trapped gas is
\begin{equation}
T_\mathrm{F} = \frac{\hbar \omega}{k_\mathrm{B}}  (6 N)^{1/3},
\end{equation}
where $\omega$ is the mean oscillation frequency, 
and $N$ is the atom number. At
fixed $T/T_\mathrm{F}$, $N$ is proportional to $\omega^{-3}$, so
higher oscillation frequencies make possible the study of quantum degenerate
gases with smaller atom
numbers. Systems with small numbers of trapped fermions are
reminiscent of quantum dots, where small numbers of electrons are
captured in electrostatic wells. Mesoscopic {\em neutral} Fermi gas
samples are also expected to exhibit rich physical phenomena. While Coulomb
interactions dominate electron mesoscopics, the relatively weaker
interaction energies of neutral Fermi systems facilitate the
study of purely quantum-statistical effects such as spatial shell
structure  \cite{shell}. 

Another appealing feature of $\mu$EM traps is the variety of trap
geometries which can be formed. Single traps with a high aspect ratio
($> 10^3$) could be used to study quasi-one-dimensional ensembles of
fermions   \cite{1dFermions} or of Bose-Fermi mixtures   \cite{1DBFmix}.
Two or more traps could be connected through quantum
point contacts  \cite{qpc} or tunneling barriers.

In the following sections, we describe our progress towards
realizing some of the experiments mentioned above using bosonic
$^{87}$Rb and fermionic $^{40}$K. After a description of the apparatus
in \S\ref{sec:apparatus}, we discuss its performance to date in \S\ref{sec:performance}, which includes
the first demonstration, to our knowledge, of trapping fermions in a
$\mu$EM trap. We discuss prospects for evaporative cooling towards
quantum degeneracy in both species and our experimental outlook in \S\ref{sec:future}.

\section{Experimental apparatus}
\label{sec:apparatus}

Experiments with quantum degenerate gases require {\em trapping}, {\em
cooling}, and {\em imaging} of ultra-cold atoms. In our case, trapping
is accomplished by capturing atoms from a background vapor in a
Magneto-Optical Trap (MOT), and then transferring them to a purely
magnetic trap, both within the same vacuum chamber.  Cooling is
accomplished in two steps: first, by laser cooling in the MOT and
with optical molasses; second, by forced evaporative cooling in the 
$\mu$EM magnetic trap. Imaging is either fluorescent or absorptive, using
resonant laser light and CCD cameras in both cases.

Our system is vastly simplified by the use of a $\mu$EM trap. The
evaporative cooling times realized in such traps are typically a few
seconds  \cite{Hansel,Lin,Esteve,Du}, short enough that a single vacuum chamber can
be used both to collect atoms from the background vapor and to cool
them to degeneracy with evaporative cooling. Systems with magnetic
traps created by standard coils require multiple chambers (or Zeeman slowers)
to maintain an ultra-high vacuum. Additional lasers are then typically required to move atoms between chambers (or slow them).  In a dual-species experiment such as
ours, additional lasers would be required for both species. The
single-chamber configuration possible with a $\mu$EM trap avoids the above 
complications.

The following subsections describe the essential components of the
experiment, sometimes focusing on technical choices that we believe
would be interesting to researchers in the field. For a more general
audience, this description can serve as a specific example of a
degenerate gas experiment that is structurally similar to many other
existing experiments. We also refer the reader to review articles of
experimental techniques \cite{ExpReview}.

\subsection{Vacuum system}
\label{sec:vacuum}

The required quality of the vacuum environment of the trap 
is in general dictated by the
desired lifetime of the trapped atoms. A collision with a background
atom or molecule (thermalized with the walls of the chamber at room temperature)
can eject an atom from the trap, since the energy
transferred is generally much higher than the trap depth (between 1\,K and 1\,mK). 
For a ten second $1/e$ lifetime for rubidium, for instance, the partial pressure of hydrogen
must be no higher than $3 \times 10^{-9}$\,mbar \cite{lifetime}. A vacuum environment also serves
to thermally isolate the cold gases from ambient temperature, since
convective and conductive heat transfer are eliminated by working in a vacuum \cite{radiative}.

Our vacuum system consists of a Pyrex cell, 75\,mm $\times$ 75\,mm  $\times$
165\,mm, connected to an ion pump and a titanium sublimation pump
through a 6 inch conflat stainless steel cube. A turbo pump is connected to the system during pump-down and bake-out, but sealed off
using a gate valve during normal operation \cite{VacuumNotes}.  The
entire system is supported on an optical table in such a way that the glass
cell is closest to the table. Magnetic trap (see \S\ref{sec:magtrap}) lifetimes as long
as 9\,s and MOT lifetimes as long as 24\,s have been observed,
suggesting that the pressure in the system is at or below $4 \times
10^{-9}$\,mbar.

\subsection{Sources}
\label{sec:sources}

Alkali atoms -- both rubidium and potassium -- are selectively
released into the vacuum chamber using dispenser sources \cite{dispensers}.
These devices use a reaction between an alkali salt
and a reducing agent to produce free alkali atoms and a byproduct. 
The reaction is activated by resistive heating of the nichrome envelope 
in which the reactants are housed.

Dispensers are commercially available \cite{SAES} for both rubidium
and potassium in their natural isotopic abundance. While the
natural abundance of $^{87}$Rb (27.83\%) is acceptably high, the
natural abundance of $^{40}$K is only 0.012\%. We use a commercial rubidium dispenser but have constructed a potassium
dispenser using an isotope-enriched potassium salt. Our dispenser was
modeled after that of Ref.~\onlinecite{K40dispenser}, with the addition of a fine
nickel mesh to contain the reactants. The device uses
the following heat-enhanced reaction to supply potassium:
\[ 2\,\mathrm{KCl} + \mathrm{Ca} \rightarrow 2\mathrm{K} + \mathrm{CaCl}_2. \]
Using hot wire tests, we estimate that our isotope-enriched dispenser can produce
100 $\mu$g of fermionic potassium.
Given our typical $^{40}$K MOT size of $10^7$ atoms (discussed in \S\ref{Recent_Results}), it is reasonable
to expect the dispenser lifetime to be on the order of years, as has
been observed in other degenerate gas experiments.

Although
some atoms dispensed by our sources are trapped in the MOT, most are
deposited on the walls of the glass cell or captured by the vacuum
pumps. We observe that $^{87}$Rb atoms can be ``recycled'' from the
glass walls using light-induced atom desorption (LIAD)  \cite{LIAD,LIADfrequ}.
Whereas previous
observations of this effect in cold atom experiments have used broadband white light
sources \cite{LIAD_MOT,Hansel}, we observe a wavelength dependence of the  $^{87}$Rb desorption efficiency\cite{potassiumLIAD}.

With the dispensers off, we measured the number of $^{87}$Rb atoms
captured by the MOT as a function of the center wavelength of various
weak ($\sim$0.5\,mW) LED sources of desorbing light. Wavelengths
greater than 500\,nm had no visible effect, whereas wavelengths in the
400\,nm range boosted the MOT loading rate by a factor of
five. This is consistent with observations in paraffin-coated vapor
cells \cite{LIADfrequ}, although it is unknown if the LIAD mechanism is
identical.  Encouraged by these preliminary tests, we built a 140\,mW
source at a center wavelength of 405\,nm. Desorption using this source
allows us to load $10^9$ $^{87}$Rb atoms into the MOT without turning
on the dispensers \cite{DApc}, which reduces the background
pressure and increases the trap lifetime. The atom number we capture
with LIAD in a UHV cell is 100 times larger than without LIAD, and over
10 times larger than that reported in
previous experiments \cite{Hansel,Du}, where higher power white light sources
were used. This comparison may not be an
accurate measure of the source efficiency, since our MOT uses
much larger trapping beams. Nonetheless, LIAD allows us to load large MOTs while maintaining a better vacuum for longer magnetic trap lifetimes.

\subsection{Laser system}
\label{sec:lasers}

\begin{figure}[t!]
\centerline{\includegraphics[width=3.25in]{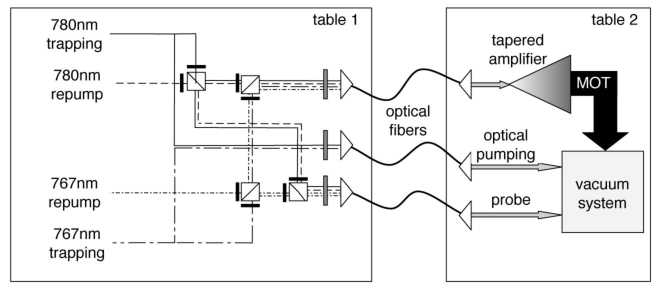}}
\caption{A schematic showing the splitting and combining of the the various optical frequencies
according to their function and desired path through the vacuum
chamber. Polarizing beam splitter cubes, broadband waveplates (black),
and dichroic wave plates (grey) are shown.  Single-mode optical fibers
send light to the vacuum system. One of these fibers injects the
tapered amplifier, which produces the MOT beams.  The other fiber
paths are aligned directly into the pyrex vacuum cell. Additional 
frequency modifications by acousto-optic modulators are not shown.}
\label{fig:lasers}
\end{figure}

Near-resonant lasers are used for laser cooling and trapping, optical
pumping, and imaging.  These applications require laser linewidths
that are small compared to the natural atomic linewidth ($\Gamma
\approx 2 \pi \times 6$\,MHz for the cycling transitions in $^{40}$K
and $^{87}$Rb) and sufficient power to saturate the cooling transition
in our MOT. These constraints are met by locking low-power
grating-stabilized diode lasers, injecting into free-running seed
lasers, passing the light through single-mode optical fibers, and
where necessary, injecting into a tapered amplifier as a second stage
of amplification, as shown in Figure~\ref{fig:lasers}.
The fiber optics produce clean spatial modes for imaging beams, and
decouple the alignment of the beams around the vacuum system from other, less critical
alignments.

Four master lasers are used: two for the $^{87}$Rb D$_2$
line at 780.246\,nm, and two for the $^{40}$K D$_2$ line at
766.702\,nm. Of each pair, one laser is tuned to the cycling transition
(i.e., from the $5\mathrm{S}_{1/2}$ $F=2$ ground state to the $5\mathrm{P}_{3/2}$$F'=3$ excited state for
$^{87}$Rb, and from the $4\mathrm{S}_{1/2}$ $F=9/2$ ground state to the $4\mathrm{P}_{3/2}$ excited state
manifold for $^{40}$K) and the other laser to the repumping transition
(out of the $F=1$ ground state for $^{87}$Rb and out of the $F=7/2$
ground state for $^{40}$K). The ground state hyperfine splittings are
6.835\,GHz and 1.286\,GHz for $^{87}$Rb and $^{40}$K respectively,
large enough that separate master lasers are convenient.

These resonant wavelengths are an ideal combination for a two-species
experiment \cite{Goldwin}.  They are close enough that they can share the same
mirrors, fibers, coatings, and tapered amplifier (whose bandwidth is 23\,nm), but far enough apart
that dichroic waveplates can selectively manipulate the polarization
of one wavelength without affecting the other, such that 100\%
combining efficiency can be achieved.

Each master is locked using saturated absorption spectroscopy. A
modulation transfer scheme is used, in which the frequency of the pump
is dithered (in our case, at 100\,kHz) and the resultant amplitude modulation of the probe
is used to lock to a peak maximum. The output of a lock circuit is
sent both to the piezo voltage and to the current of the master laser,
to achieve a lock bandwidth of over 5\,kHz. Using this method, we observe a short term
stability of 300\,kHz.
Three of these master lasers (all but the Rb repump) inject room-temperature free-running diode lasers to boost their power.

Figure~\ref{fig:lasers} shows the mixing of various frequencies into
the fiber optics that transport them to the vacuum system.  The
majority of the laser power is injected into the tapered amplifier,
producing 0.65\,W of output power after spatial filtering, 
which is split into six beams for the
MOT. A second fiber is used for imaging, which is described in more
detail in \S\ref{sec:imaging}. A third fiber is used for optical
pumping into a specific magnetic sublevel to maximize efficiency of
loading from the MOT (which traps all states) into the magnetic trap
(which traps only specific sublevels). In all cases, dichroic
waveplates \cite{dichroicNotes} are used to produce a single
polarization of light.

\subsection{Magnetic trap}
\label{sec:magtrap}
\begin{figure}[b!]
\centerline{\includegraphics[width=3.375in]{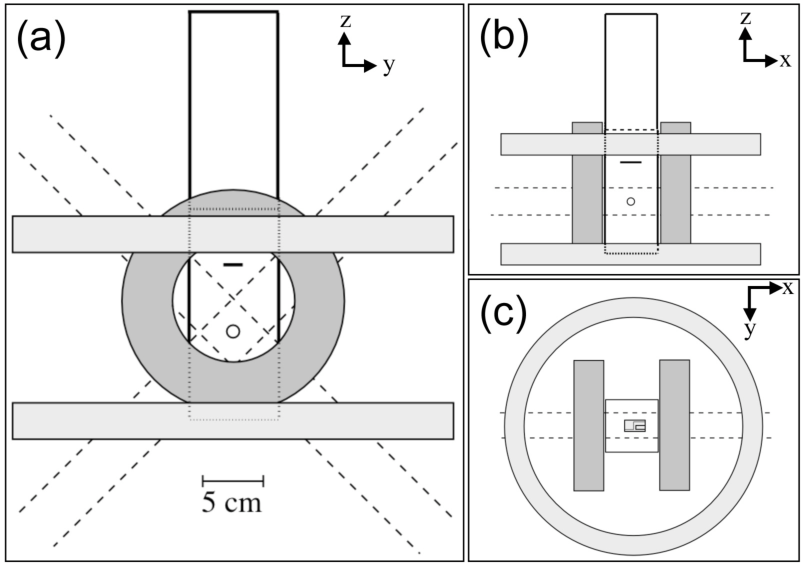} }
\caption{Glass cell, MOT coils, Transfer coils, atom chip, and MOT
beams as seen along {\bf(a)} the x axis, {\bf(b)} the y axis; and
{\bf(c)} the z axis. Coils (shaded) are positioned around
the glass vacuum cell (solid lines). Diagonal ($\hat{y} \pm \hat{z}$)
MOT beams are shown (dashed) only in (a); horizontal ($\hat{x}$) MOT
beams are shown in both (b) and (c). The atom chip is seen face-on in
(c) and indicated with a thick black line in (a) and (b).}
\label{fig:trap}
\end{figure}
\begin{figure*}[t!]
\centerline{\includegraphics[width=5in.]{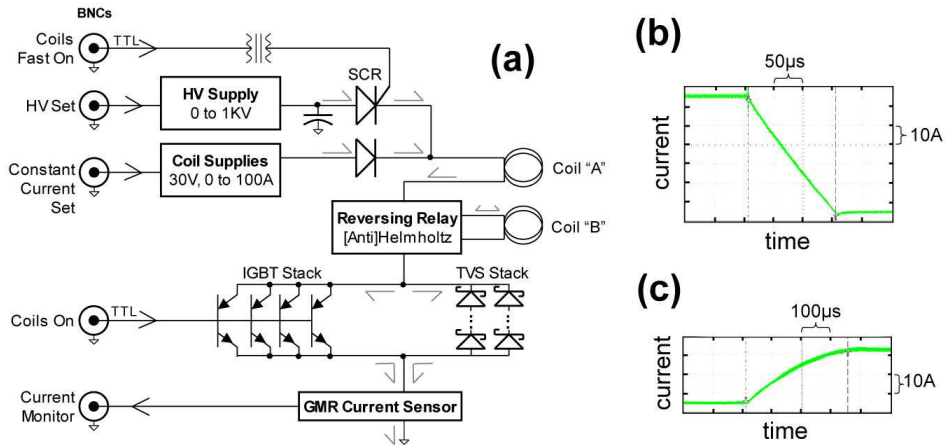}}
\caption{{\bf (a)} Magnetic trap switch for the MOT coils. The primary
switch is a stack of Insulated Gate Bipolar junction Transistors
(IGBT's). Turn-on is accelerated with a high voltage (HV) charge in a
capacitor, whose discharge is triggered using a Silicon Controlled
Rectifier (SCR). Turn-off is accelerated with transient voltage
suppressors (TVS's). Current is sensed using a giant magneto-resistive
(GMR) current sensor. Arrows indicate the direction of controls and
monitors; half-arrows indicate coil currents.
{\bf (b)} Turn-off performance: current falls from 60\,A to zero in
150\,$\mu$s.  {\bf (c)} Turn-on performance: rise of current from zero
to 26.5\,A in 350\,$\mu$s.  }
\label{fig:magomatic}
\end{figure*}

Two pairs of circular coils are used to create the magnetic field for
the MOT, for the first magnetic trap, and for the movement of the
magnetically trapped atoms from the first magnetic trap (formed at the
MOT) to the chip trap 5\,cm away. As shown in Figure~\ref{fig:trap}, all four
coils are mounted outside the glass cell. The inner pair of coils (the
``MOT coils'') create the linear quadrupole fields for the MOT and
magnetic trap.  Both MOT coils consist of 100 turns of insulated,
hollow-core \cite{wireNotes} copper wire, wound to form a square
cross-section with an inner (outer) diameter of 10 cm (18.4 cm).  The coils
are mounted with an inner separation of 8.4 cm. The outer coils (the
``Transfer coils'') carry equal and parallel currents to provide a
nearly uniform field in the $z$ direction.  They are made of the same
hollow-core copper wire, but have 49 turns each and inner (outer)
diameters of 28\,cm (36.4 cm).  The Transfer coils are mounted with an
inner separation of 11.4\,cm.  Applying the uniform field of the
Transfer coils to the quadrupole field of the MOT coils shifts the
location of the $B=0$ trap center along the $z$ axis (vertical).  The
MOT laser beams are aligned so that they intersect at $z = -2.5$\,cm
from the center of the MOT coils.  Together, these coils create
and move the trap center between the location of the MOT and the surface
of the chip, several centimeters away.

The advantage of this ``off-center'' MOT is power efficiency.  A
strong quadrupole trap can only be formed within roughly half a coil
radius of the center of a pair of anti-Helmholtz coils.
Had we formed the MOT at the
center of the coils and transported the atoms the same
distance to the chip, the radius of
the coils would have to have been doubled.
Maintaining the same magnetic field gradient in these larger
coils would require an order of magnitude more electrical power.

As built, the power consumption for the primary coils is 0.8\,kW at
60\,A, which generates a gradient of 94\,G/cm along the $\hat{x}$-axis. 
Optical access is 
sufficient for six 5-cm-diameter MOT beams. 
Using the Transfer coils, cold atoms have been transported over
5.5\,cm.

\subsection{Trap switch}

Several phases of the experiment require fast switching of the
magnetic trap coils. In order to load atoms from the MOT into the
magnetic trap, the coils must be turned off quickly to create a
zero-field condition for optical molasses, and then turned on quickly
to high current for efficient loading of the magnetic trap. Fast
turn-off is also necessary when imaging the atoms in order to avoid
heating, distortion of the cloud, and Zeeman shifts of the atomic
energy levels. The necessary time scale is 
determined by the motion of
atoms in the trap: fields should be switched in less time
than a single classical oscillation period.  A chip trap (see \S\ref{sec:chip}) can have oscillation
frequencies above 1\,kHz, requiring sub-ms switching times for its bias
fields.  At 3\,mH inductance and 60\,A of current, voltages up to 1\,kV are required to achieve sufficiently fast switching
times.

Figure~\ref{fig:magomatic}a shows a schematic of the control circuit (dubbed
``Mag-O-Matic'') for rapidly switching large currents. The
complete trap requires two such switches: one for the MOT and one for
the Transfer coil. These are independent and nearly identical; only
the MOT control circuit is shown. The difference between the two
circuits is the role of the relay: the Transfer coils are hard-wired
in series, but the relay determines the direction of current; the relay
in the MOT circuit switches the coils between Helmholtz and
anti-Helmoltz configurations.

When the coils are operating in steady state, the IGBT (Insulated Gate
Bipolar junction Transistor) stack is on and the coil supplies are
running in constant current mode. The current flows through the
blocking diode, the coils, the IGBT stack, and a giant
magneto-resistive (GMR) current sensor.  The sensor is used for
quantitative monitoring only.

The coils are turned off by turning the IGBT stack off.  The
counter-EMF (or ``flyback'') caused by the inductance of the coils
produces a high voltage spike which is clamped by the transient
voltage suppressors (TVS) at about 940\,V.  The coil current is
dissipated at 0.4\,A/$\mu$s, as shown in Figure~\ref{fig:magomatic}b.

There are two ways to turn on the coil current.  If the IGBT stack
alone is turned on, the current will rise asymptotically to the
steady-state value in $\sim$20\,ms. The other turn-on method (the
``fast on'') works as follows.  At any time before the fast turn-on is
needed, the high voltage (HV) supply is used to charge up the HV
capacitor.  Fast turn-on is triggered by first turning on the IGBT
stack, and then triggering the HV Silicon Controlled Rectifier
(SCR). In the few microseconds between these two events, no
significant current flows due to the coil inductance.  Triggering the
SCR creates an LC parallel resonant circuit with the coil inductance
and the HV capacitor. During the first quarter-cycle of the resonance,
roughly 350\,$\mu$s, the capacitor transfers its charge to the coil
(see Figure~\ref{fig:magomatic}c).  The charge on the capacitor is
chosen such that after this quarter-cycle the coil is at its steady
state voltage, and the blocking diode of the supplies will conduct to
provide the steady state current.  The SCR turns off when the current 
flowing through it falls to zero.

\subsection{Atom chip}
\label{sec:chip}

\begin{figure}[b!]
\centerline{\includegraphics[width=5cm]{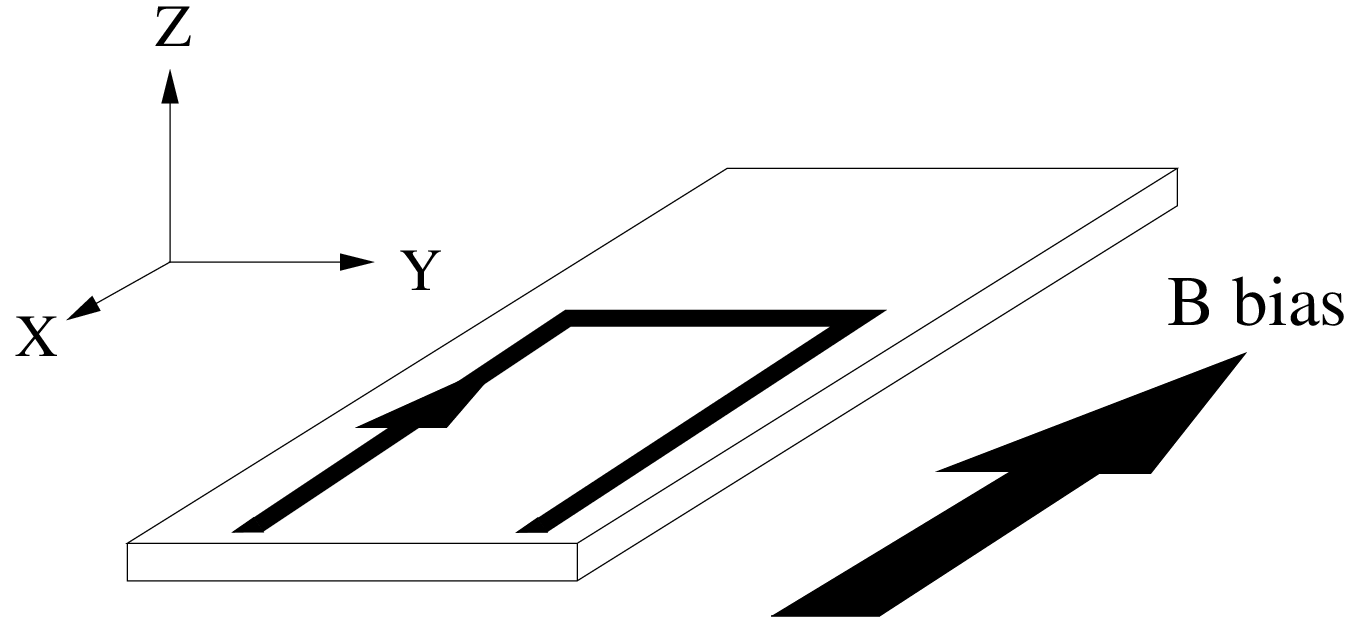}}
\centerline{\includegraphics[width=5cm]{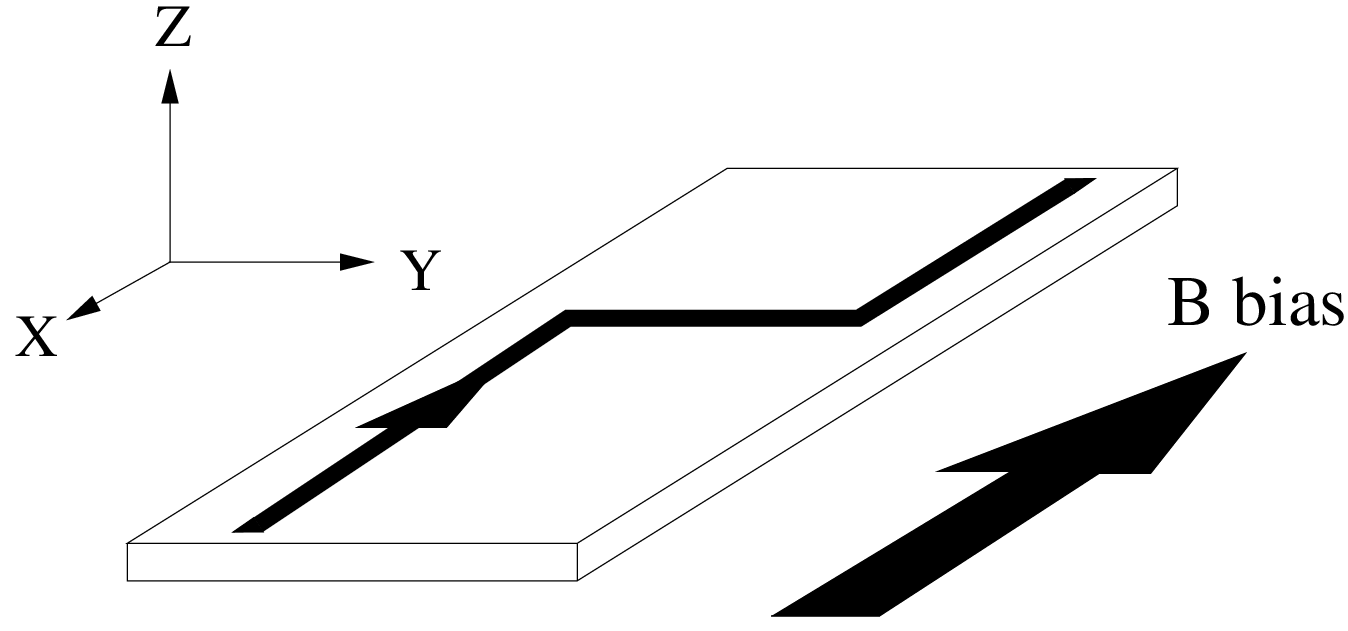}}
\caption{Schematic diagrams of the U (upper) and Z (lower) wire trap
configurations.  The arrows on the chip surfaces indicate the
direction of DC current flow; a uniform bias field is applied in the
direction indicated. These two trapping wires are fabricated on the
same chip, but shown separately here for clarity. }
\label{fig:chip}
\end{figure}

Although the external magnetic trap described in \S\ref{sec:magtrap}
can capture and hold the atoms, efficient evaporative cooling requires
a collision rate that is several hundred times greater than the loss
rate from the trap \cite{evaporation}. In our single-chamber vapor
cell, the loss rate is roughly 0.2\,s$^{-1}$.
As a result, the initial collision rate must be greater than
$10^2$\,s$^{-1}$, which (for the number and temperature typically
captured by a MOT) requires a more tightly confining trap than our
external coils can provide.

A $\mu$EM trap (or ``atom chip'') can provide much
stronger confinement -- while using smaller currents -- than a trap
produced by conventional coils \cite{Weinstein}. To understand this,
consider a cylindrical wire of diameter $D$ carrying current $I$. The
field gradient at a distance $r>D/2$ from the wire is
\begin{equation}
B' = -\frac {\mu_0}{2 \pi} \frac{I}{r^2}, 
\end{equation}
where $\mu_0$ is the permittivity of free space. Decreasing the wire radius
decreases the minimum $r$ but also decreases the current
capacity \cite{groth} $I$ as $D^{3/2}$, and thus the maximum gradient
scales as $D^{-1/2}$. Reducing $D$ from millimeter-scale to
micrometer-scale therefore improves the magnetic field gradient by
several orders of magnitude. This higher gradient provides tighter
confinement, which increases the collision rate and accelerates
the thermalization necessary for evaporative cooling.

The atom chip we are using for this experiment is identical to the one
described in Ref.~\onlinecite{Esteve}. Two types of traps are used: a
``U-trap'' and a ``Z-trap''. These configurations, along with the
external field necessary to form a trap near the wire, are shown in
Figure~\ref{fig:chip}.

The U-trap consists of a 420-$\mu$m-wide, 7-$\mu$m-high gold wire in
the shape of the letter `U'. The
base of the U is 2.0\,mm long. When a bias field is applied parallel to
the plane of the U and perpendicular to its base, a three-dimensional
quadrupole trap is formed above the base of the wire. Using a current
of 5.0\,A and a bias field of 20\,G,  a quadrupole trap is formed
430\,$\mu$m from the surface, with a trap depth of 1.3\,mK, and a mean
gradient of 200\,G/cm.

The Z-trap consists of a 60-$\mu$m-wide, 7-$\mu$m-high gold wire in
the shape of the letter `Z', as shown in Figure\,\ref{fig:chip}. The
middle segment of the Z is 2.85\,mm long. When a bias field is applied
to this wire, a Ioffe-Pritchard-type trap (i.e., a magnetic trap whose
minimum field is nonzero) is formed above the surface. Using a current
of 2.0\,A and a bias field of 40\,G, the minimum is 100\,$\mu$m from
the surface, with an oscillation frequency of 6.5\,kHz in the
transverse direction and 13\,Hz in the longitudinal
direction. These calculations are for $^{40}$K, and assume a 1\,G ``Ioffe field'' applied in
the $\hat{y}$ direction, to increase the minimum field of the trap.

\subsection{Sequencing}
\label{sec:sequencing}

The experiment operates in a cycle, starting with the collection of
hot atoms and progressively compressing and cooling the cloud to
higher phase space density. Coordination of each stage of the
experiment is accomplished using a dedicated real-time system sequencer \cite{adwinNotes}. The sequencer operates with an 80\,MHz
clock rate, and has 128\,MB of onboard memory, allowing for the generation of
complex waveforms. Twenty-four analog outputs and thirty-two digital
outputs are updated as quickly as once every 10\,$\mu$s. These outputs
are buffered (as described below) and sent to various parts of the
experiment: shutters, frequency modulators, power supplies, cameras,
and the magnetic switch. A standard desktop computer programs the
sequencer over an ethernet connection.

\begin{figure}[t]
\centerline{\includegraphics[width=3.375in]{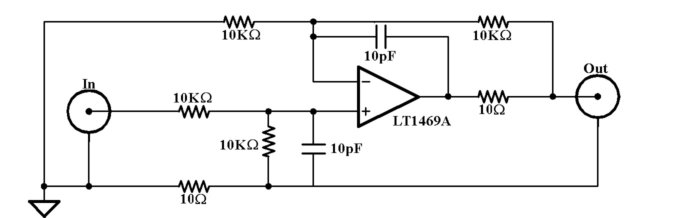} }
\caption{Schematic diagram of one channel of the analog buffer.}
\label{fig:analogbuffer}
\end{figure}

In order to protect the sequencer, drive low impedance loads, and
prevent ground loops, buffers are used for all channels of the
sequencer. Galvanic isolation of the digital channels is accomplished
by using a GMR-based device \cite{dbNotes}. The schematic used for the
analog buffer is shown in Figure~\ref{fig:analogbuffer}. Outputs from
the buffer have ``soft'' grounding: the 10~$\Omega$ between the source
ground and the buffer output return is low enough to carry any load
current yet high enough to prevent ground loops.  A variation on a
differential input amplifier corrects for the output return's offset
and noise.

\subsection{Evaporative cooling system}
\label{sec:rf}

Laser cooling of atoms is restricted to low densities because light
must be able to escape from the cloud to carry away entropy.  In
essence, this constraint limits most laser cooling schemes to phase
space densities below $10^{-4}$. Further cooling can be accomplished
in a purely magnetic trap using evaporative cooling, in which the
hottest atoms are selectively removed from the trap, each carrying
away more than the average energy of a trapped atom.  After
rethermalization, the remaining ensemble has a lower
temperature. Atoms are selectively removed using an electromagnetic
field that couples only to atoms located at higher potential energy
surfaces in the trap. Generally, the field is in the radio frequency
(RF) range. The RF field spin-flips atoms to a magnetically
anti-trapped state, and they are ejected from the trapping volume.

While many quantum gas experiments use standard commercial function generators, 
such devices typically have dwell times and switching times of
10\,ms (or greater), rendering them unable to sweep quickly enough for
an evaporation that may last only one second. Instead we generate
the RF field using Direct Digital Synthesis (DDS), which has
sufficient speed and precision for our application, at a fraction of
the cost of a complete high-frequency function generator. DDS creates
a sinusoidal waveform point by point, which is then sent through a
digital to analog converter \cite{ADNotes}. 
Using dividers implemented in digital logic, any frequency
from $\mu$Hz up to the Nyquist frequency (146\,MHz in our case) can be
generated with 48-bit resolution.  Linear sweeps can be produced by
specifying beginning and ending frequency, number of frequency steps,
and step time. Since no phase-locked loop is involved, the sweep rate is
limited only by the digital clock cycle. Exponential sweeps are
approximated piecewise by a series of linear sweeps.

The frequency and amplitude of the DDS are programmed using three
digital lines from the sequencer. The RF output signal is amplified
and fed into an auxiliary wire on the chip, which acts as an
``antenna''.  Since the antenna is less than a millimeter away from
the atoms, little power is required to perform RF evaporation.

\subsection{Imaging}
\label{sec:imaging}

We obtain information about the atom cloud with fluorescence and
absorption imaging.
Fluorescence imaging is accomplished by observing the light
of a resonant laser beam scattered by an atom cloud. In absorption imaging, a
resonant beam is passed through the cloud, and imaged on a CCD camera.
The atoms absorb a portion of the laser beam, leaving behind a shadow
of the atom cloud. Since the optical density of the cloud is
proportional to the three-dimensional density integrated along the
line of observation, the shadow gives direct information about the
spatial distribution of the atom cloud.

At the chip, we also detect the cold atomic cloud by
generating a one-dimensional Magneto-Optical Trap (1D MOT). 
A retro-reflected probe laser parallel
to the chip surface is directed onto the magnetic field minimum of the
chip trap, where the atoms are located. When the probe laser is
red-detuned from resonance, the cold atomic cloud scatters light,
while being confined by the magneto-optical forces in
the direction of the probe propagation.  
The longer interrogation time in the 1D MOT allows for 
a higher signal-to-noise detection than simple fluorescence imaging.
The method is useful for
measuring atom number, but it does not image the cloud, since the atoms
are pushed towards the magnetic center of the 1D MOT. 
(In fact, this technique is a useful way to
determine the locations of magnetic field minima above the chip.) 

Four different imaging systems are used in the experiment: one looking
from below along the $\hat{z}$-axis, two to image clouds at the chip
from the $\hat{x}$ and $\hat{y}$ directions, and one to image the MOT
and first magnetic trap.  Each imaging system
makes use of two lenses in order to form an image of the cloud on the
CCD of the camera. The two chip imaging systems are
built to observe small clouds, and use a matched set of achromats with
numerical apertures of 0.33 for the $\hat{x}$-axis system and 0.25 for
the $\hat{y}$-axis system.

We use monochrome CCD cameras with $640\times480$ pixels, $7.4$\,$\mu$m
$\times7.4$\,$\mu$m each, with a variable-gain 10-bit intensity
greyscale \cite{camNotes}. The acquisition duration is electronically
shuttered and can be varied in steps of 50\,$\mu$s.  Communication with
imaging software over an IEEE1394 (``Firewire'') communication
standard obviates the need for a dedicated imaging board.

\section{Trapping of $^{87}$Rb and $^{40}$K on a chip}
\label{sec:performance}

Even though the experimental apparatus described in \S\ref{sec:apparatus} 
is still in the final stages of
construction and testing, we have successfully
cooled and trapped atoms on the chip.  
We describe (in \S\ref{Rb87_Sequence_subsection}) the
sequence we use for loading cold $^{87}$Rb, 
and (in \S\ref{Recent_Results}) a recent result in which we
have used a variation on the sequence to simultaneously load
$^{87}$Rb and $^{40}$K into a chip trap.

\subsection{Experimental sequence for $^{87}$Rb} 
\label{Rb87_Sequence_subsection}

From the background vapor produced by a rubidium dispenser, we trap
and cool up to $10^9$ $^{87}$Rb atoms in a magneto-optical trap (MOT).
The trapping beams 
each have an intensity of about 2.5\,mW/cm$^2$, and are detuned $\delta
= -4.5\Gamma$ from the 5S$_{1/2}$F=2 $\rightarrow$ 5P$_{3/2}$F=3 cycling
transition. The repumper is 0.8$\%$ of the trapping intensity. The
MOT coils provide a gradient of 9\,G/cm along the strong
axis. We cool the atoms further by rapidly turning off
the current in the anti-Helmholtz coils and detuning the trapping
beams to $\delta = -9\Gamma$ for 9\,ms to create an optical
molasses. After this second cooling stage, we measure a temperature of
T$\leq 30 \mu$K and a phase space density $\rho \approx 3 \times 10^{-6}$.
\begin{figure}[t!]
\centerline{\includegraphics[width=3.25in]{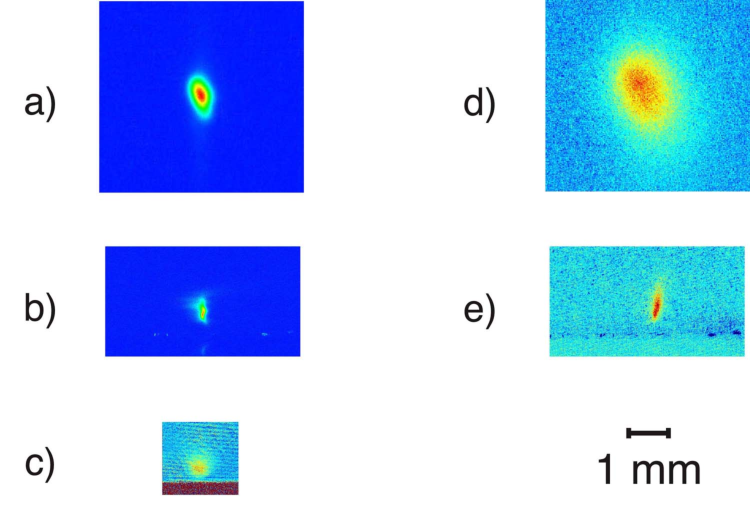}}
\caption{Images of $^{87}$Rb (left column) and $^{40}$K (right column)
clouds in the  magneto-opitcal trap and at the chip:
{\bf (a)} $^{87}$Rb MOT;
{\bf (b)} $^{87}$Rb after $\mu$EM trapping, imaged using a 1D MOT at the chip;
{\bf (c)} Absorption image of  $^{87}$Rb after release from the $\mu$EM trap;
{\bf (d)} $^{40}$K MOT;
{\bf (e)} $^{40}$K  after $\mu$EM trapping, imaged using a 1D MOT at the chip.
More detail about atom numbers and imaging techniques are
given in the text. All images use the same spatial scale, which is
indicated.} 
\label{image}
\end{figure}

After the molasses stage, the atoms are recaptured in a purely magnetic
trap with a vertical gradient of 21\,G/cm. 
This magnetic field gradient is strong enough to
trap atoms in the $\textrm{F}=2, \textrm{m}_F=2$ hyperfine ground
state against the force of gravity, while atoms in
lower magnetic substates remain untrapped. Once trapped, the atoms are
compressed adiabatically for transport to the chip with a vertical
gradient of 47\,G/cm (94 G/cm on the strong horizontal axis of the
trap).  Typically, we load $15\%$ of the atoms ($1.5 \times 10^8$
atoms) into the quadrupole trap, and after compression, they have a
temperature of T $\approx$100$\,\mu\textrm{K}$ and a phase space
density of $\rho\approx10^{-7}$. 

The efficiency of the quadrupole trap loading is improved by the
insertion of a short optical pumping step immediately after the optical
molasses stage. We turn on a uniform bias magnetic field to provide a
quantization axis for the atoms and then apply along this same axis a
200\,$\mu$s laser pulse tuned to the 5S$_{1/2}$F=2 $\rightarrow$
5P$_{3/2}$F=2 transition to optically pump the atoms. By using
$\sigma+$ polarized light, the atoms are pumped into the
$F=2, m_F=2$ hyperfine ground state.  This step
typically improves the loading efficiency by a factor of two, while
leaving the size and temperature of the atomic cloud
essentially unchanged.

We transport the atoms in the quadrupole trap up to the chip by
changing the current in the Transfer coils.  With a magnetic field
gradient of 47\,G/cm in the vertical direction, we move the atoms
from the MOT position to the chip by smoothly varying the vertical ($+\hat{z}$)
bias magnetic field (produced by the Transfer coils) from $-84$\,G
to $+121$\,G in about 600\,ms. The atoms do not experience any
significant loss or heating during the transport. Horizontal ($\hat{x}$ and $\hat{y}$) bias
magnetic fields provide transverse positioning to align the cloud to the center
of the $\mu$EM trap.

Once at the chip, the atoms are loaded into the
U-wire trap by first increasing the current in the U-wire, and then
simultaneously decompressing the quadrupole trap while increasing the
transverse horizontal bias magnetic field for the chip trap. The
loading process takes 300\,ms. After loading the U-wire trap, we find
that the trap contains $1 \times 10^7$ atoms with a temperature of
T=200\,$\mu\textrm{K}$ and a phase space density of $\rho\approx10^{-7}$.

This entire sequence can be initiated at a lower vapor pressure by turning
off the dispensers and using LIAD, as is described in \S\ref{sec:sources}. 
Figure~\ref{image} shows
images of light-desorbed $^{87}$Rb in the MOT (Fig.~\ref{image}a) and 
after release from the chip trap (Fig.~\ref{image}b,c).

\subsection{Simultaneous loading of $^{87}$Rb and $^{40}$K}
\label{Recent_Results} 

Recently, we have adapted the above
sequence, developed for $^{87}$Rb, to include $^{40}$K. We report two
technical milestones: (1) trapping of a cold fermionic atomic species in a chip trap, and (2)
simultaneous trapping of two elements, in this case a bosonic isotope and
a fermionic isotope, in a chip trap.

In order to load $^{40}$K onto the chip, 
we make the following changes to the sequence.
We turn on the $^{40}$K-enriched dispenser and potassium lasers to load a MOT.
Then, instead of using an optical molasses cooling stage, we
compress the MOT for 100\,ms with an anti-Helmholtz gradient of
94\,G/cm. The trapping beams are then turned off and the atoms remain
trapped in the magnetic quadrupole field (which is already on at full
strength).  The rest of the sequence is as described in \S\ref{Rb87_Sequence_subsection}: we transport 
the cold atoms to the chip and transfer them into the $\mu$EM trap.

Using this sequence, $^{40}$K is loaded onto the chip. 
Figure~\ref{image}d shows a $^{40}$K MOT and Figure~\ref{image}e shows
a $^{40}$K cloud that was released from the $\mu$EM trap and imaged using the 
1D-chip-MOT fluorescence detection scheme described in \S\ref{sec:imaging}.

Turning on the LIAD source and $^{87}$Rb lasers, we observe that {\em both }
species are loaded onto the chip, although $^{87}$Rb is not loaded as efficiently as in 
the original sequence. By selecting a probe beam
at 780\,nm or 767\,nm, we can detect $^{87}$Rb or $^{40}$K atoms,
independently. We observe simultaneous
trapping of $2\times10^5$ $^{87}$Rb atoms and $6\times10^4$ $^{40}$K
atoms in the U-wire trap.

\subsection{Discussion and future work}

While the above sequences are successful, they are not optimal.
The transfer efficiency ($\sim10\%$) from the external
quadrupole trap to the U-wire $\mu$EM trap is not as high as 
has been previously observed in similar transfer sequences \cite{Lin}. 
It is possible that we are limited by the trap depth of the $\mu$EM trap.

The sequence used to load $^{40}$K is preliminary and has not been studied carefully.
Unlike the sequence described in \S\ref{Rb87_Sequence_subsection}, the method described in \S\ref{Recent_Results} to load the magnetic trap 
will trap more than one m$_\mathrm{F}$ level
in each species. A spin mixture in the magnetic trap is susceptible to spin-exchange losses. 
The range of magnetic sublevels can be reduced with an optical pumping step.

Future work includes the evaporative cooling of $^{87}$Rb and
sympathetic cooling of $^{40}$K, both to quantum degeneracy. 
For quantum degeneracy, atoms must be loaded
into the Z-wire trap, whose Ioffe-Pritchard field configuration avoids
spin-flip losses. Improvements in loading the chip will be
necessary to achieve the correct starting condition for evaporative
cooling to degeneracy.

Several additions to our apparatus are possible.
Manipulation of the atoms with microwaves will enable the simultaneous
evaporation of $^{40}$K and $^{87}$Rb, which has been shown to improve
efficiency in other dual species experiments \cite{DualEvap}. Far
off-resonant laser light can be used to form a trap using the induced
electric dipole of the atom.  Such an optical trap is useful for the
study of multiple spin states, and for accessing Feshbach-Verhaar
resonances \cite{Feshbach}. A next-generation microfabricated trap will be built to
reduce light scattering off the chip, and to address known problems
with the techniques used to fabricate our chip \cite{Esteve}. Finally,
a higher quality camera will improve our imaging sensitivity.

\section{Conclusion}
\label{sec:future}

In conclusion, we have demonstrated the trapping of $^{40}$K and of
$^{87}$Rb on a chip. To our knowledge, this is the first time fermions
have been confined by a $\mu$EM trap, and the first time two elements
have been simultaneously loaded into a $\mu$EM trap. The apparatus used to
achieve this has been described in detail, indicating some technical
choices that may interest researchers in the field.

The high aspect ratio of these traps may be useful for studies of
quasi-one-dimensional ensembles. In the simplest case, the aspect
ratio is controlled by the ratio
of the wire current to the bias field: the closer the trap is to
the surface, the higher its aspect ratio. The 500:1 aspect ratio of
the Z-trap in the example given in \S\ref{sec:chip} seems well suited
to quasi-one-dimensional experiments. A complication is the trap
``roughness,'' which has been investigated in
Refs.~\onlinecite{surfaces,Esteve}. This roughness may spoil the use of some
wire traps as simple elongated traps, but may also allow for the study
of quantum gases in potentials with frozen-in
randomness \cite{lukinRandom}.

\bigskip
{\it Note added in proof:} Since the original submission of this paper, we have been able to load simultaneously $2 \times 10^7$ $^{87}$Rb atoms with a phase space density of $3 \times 10^{-6}$, along with  $2 \times 10^5$  $^{40}$K atoms in the $|F=9/2,mF=9/2>$ state into the Z-wire trap.

\section*{Acknowledgments}
We would like to thank Hyun Youk and Ryan Bolen for early work on this
project. We also thank Thorsten Schumm, Brian DeMarco, Vladan Vuleti\'{c}, 
Peter Kr\"{u}ger, Sebastian Hofferberth, J\"{o}rg Schmiedmayer, and Joszef Fort\'agh for
stimulating and helpful conversations. S.A. and L.L. acknowledge support from NSERC;
M.H.T.E. acknowledges support from OGS. This work was supported by CFI, OIT, NSERC, PRO, and Research Corporation.



\begin{thebibliography}{99}


\bibitem{BEC}M.\ H.\ Anderson, J.\ R.\ Ensher, M.\ R.\ Matthews, C.\
E.\ Wieman, and E.\ A.\ Cornell, \emph{Science} \textbf{269} 198 (1995);
K.\ B.\ Davis, M.-O.\ Mewes, M.\ R.\ Andrews, N.\ J.\ van Druten, D.\
S.\ Durfee, D.\ M.\ Kurn, and W.\ Ketterle, \emph{Phys.\ Rev.\ Lett.}
\textbf{75}, 3969 (1995); C.~C.~Bradley, C.\ A.\ Sackett, J.\ J.\
Tollett, and R.\ G.\ Hulet, \emph{Phys.\ Rev.\ Lett.} {\bf 75}, 1687
(1995); see also C.\ A.\ Sackett, C.\ C.\ Bradley, M.\ Welling, and R.\
G.\ Hulet, \emph{Appl.\ Phys.\ B} \textbf{B65}, 433 (1997).

\bibitem{FDG}{ B.\ DeMarco and D.\ S.\ Jin, \emph{Science} \textbf{285},
1703 (1999); 
A.\ G.\ Truscott, K.\ E.\ Strecker, W.\ I.\ McAlexander,
G.\ B.\ Partridge, and R.\ G.\ Hulet, \emph{Science} {\bf 291}, 2570 (2001); 
F.\ Schreck, L.\ Khaykovich, K.\ L.\ Corwin, G.\ Ferrari, T.\
Bourdel, J.\ Cubizolles, and C.\ Salomon, \emph{Phys.\ Rev.\ Lett.} {\bf
87}, 080403 (2001); 
S.\ R.\ Granade, M.\ E.\ Gehm, K.\ M.\ O'Hara, and J.\
E.\ Thomas, \emph{Phys. Rev. Lett.} {\bf 88}, 120405 (2002); 
Z.~Hadzibabic, C.~A.~Stan, K.~Dieckmann, S.~Gupta, M.~W.~Zwierlein, A.~G\"orlitz, and W.~Ketterle, \emph{Phys.\ Rev.\ Lett.} {\bf 88}, 160401 (2002);
G.~Roati, F.~Riboli, G.~Modugno, and M.~Inguscio, \emph{Phys.\ Rev.\ Lett.} {\bf 89}, 150403 (2002);
M.~Bartenstein,  A.~Altmeyer, S.~Riedl, S.~Jochim, C.~Chin, J.~Hecker Denschlag, and R.~Grimm, \emph{Phys.\ Rev.\ Lett.} {\bf 92}, 120401 (2004); 
M.~K\"ohl, H.~Moritz, T.~St\"oferle,  K.~G\"unter, T.~Esslinger, \emph{Phys.\ Rev.\ Lett.} {\bf 94}, 080403 (2005); 
C.~Silber, S.~Guenther, C.~Marzok, B.~Deh, Ph.~W.~Courteille, C.~Zimmermann, \mbox{cond-mat/0506217};
C.~Ospelkaus, S.~Ospelkaus, K.~Sengstock, K.~Bongs, \mbox{cond-mat/0507219}. }

\bibitem{LukinHTSC} W.\ Hofstetter, J.\ I.\ Cirac, P.\ Zoller, E.\
 Demler, and M.\ D.\ Lukin, \emph{Phys.\ Rev.\ Lett.}, \textbf{89}, 220407
 (2002).

\bibitem{Anderson} U.\ Gavish and Y.\ Castin, \emph{Phys.\ Rev.\ Lett.} {\bf 95}, 020401 (2005).

\bibitem{BEC-BCS} C.\ A.\ Regal, M.\ Greiner, and D.\ S.\ Jin,
\emph{Phys.\ Rev.\ Lett.}, \textbf{92}, 040403 (2004); M.\
Bartenstein, A.\ Altmeyer, S.\ Riedl, S.\ Jochim, C.\ Chin, J.\ Hecker
Denschlag, and R.\ Grimm, \emph{Phys.\ Rev.\ Lett.}, \textbf{92}, 120401
(2004); T.\ Bourdel, L.\ Khaykovich, J.\ Cubizolles, J.\ Zhang, F.\
Chevy, M.\ Teichmann, L.\ Tarruell, S.\ J.\ J.\ M.\ F.\ Kokkelmans,
C.\ Salomon, \emph{Phys. Rev.\ Lett.}, \textbf{93}, 050401 (2004); M.\
W.\ Zwierlein, C.\ A.\ Stan, C.\ H.\ Schunck, S.\ M.\ F.\ Raupach, A.\
J.\ Kerman, W.\ Ketterle, \emph{Phys.\ Rev.\ Lett.}, \textbf{92},
120403 (2004).

\bibitem{Hansel} W.\ H\"{a}nsel, P.\ Hommelhoff, T.\ W.\ H\"{a}nsch,
and J.\ Reichel, \emph{Nature} {\bf 413}, 498 (2001).

\bibitem{Ott} H.\ Ott, J.\ Fortagh, G.\ Schlotterbeck, A.\ Grossmann,
and C.\ Zimmermann, \emph{Phys.\ Rev.\ Lett.} {\bf 87} 230401 (2001).

\bibitem{Weinstein} J.\ D.\ Weinstein and K.\ G.\ Libbrecht,
\emph{Phys.\ Rev.\ A}, \textbf{52}, 4004 (1995).

\bibitem{atomwg}J.\ H.\ Thywissen, M.\ Olshanii, G.\ Zabow, M.\
Drndi\'{c}, K.\ S.\ Johnson, R.\ M.\ Westervelt, and M.\ Prentiss, \emph{Eur.\ Phys.\
J.\ D} {\bf 7}, 361 (1999); D.\ M\"{u}ller, D.\ Z.\ Anderson, R.\ J.\
Grow, P.\ D.\ D.\ Schwindt, and E.\ A.\ Cornell, \emph{Phys.\ Rev.\
Lett.} {\bf 83}, 5194 (1999); N.\ H.\ Dekker, C.\ S.\ Lee, V.\ Lorent,
J.\ H.\ Thywissen, S.\ P.\ Smith, M.\ Drndi\'{c}, R.\ M.\
Westervelt and M.\ Prentiss, \emph{Phys.\ Rev.\ Lett.} {\bf 84}, 1124 (2000).

\bibitem{ReichelMotor} W.\ H\"ansel, J.\ Reichel, P.\ Hommelhoff, and T.\
W.\ H\"ansch, \emph{Phys.\ Rev.\ Lett.}, \textbf{86}, 608 (2001).

\bibitem{Folman}R.\ Folman, P.\ Kr\"{u}ger, D.\ Cassettari, B.\
Hessmo, T.\ Maier, and J.\ Schmiedmayer, \emph{Phys.\ Rev.\ Lett.}
{\bf 84}, 4749 (2000).

\bibitem{surfaces} J. Fort\'agh, H. Ott, S. Kraft, A.\ G\"unther, and C. Zimmermann,
\emph{Phys. Rev. A} {\bf 66}, 041604(R) (2002);
ÊM.~P.~A.~Jones, C.~J.~Vale, D.~Sahagun,
B.~V.~Hall, and E.~A.~Hinds \emph{Phys.~Rev.~Lett.} {\bf 91}, 080401
(2003); J.\ M.\ McGuirk, D.\ M.\ Harber, J.\ M.\ Obrecht,
and E.\ A.\ Cornell \emph{Phys.\ Rev.\ A} {\bf 69}, 062905 (2004); 

\bibitem{Lin}{Y.
Lin, I.~Teper, C.~Chin, and V.~Vuleti\'{c} \emph{Phys.~Rev.~Lett.}
{\bf 92}, 050404 (2004).}

\bibitem{shell}J.\ Schneider and H.\ Wallis, \emph{Phys.\ Rev.\ A}
 {\bf 57}, 1253 (1998).

\bibitem{1dFermions}B.\ E.\ Granger and D.\ Blume,
\emph{Phys. Rev. Lett.} {\bf 92}, 133202 (2004).

\bibitem{1DBFmix}{K.\ Das, \emph{Phys. Rev. Lett.} {\bf 90}, 170403 (2003);
M.\ A.\ Cazalilla and A.\ F.\ Ho, \emph{Phys. Rev. Lett.} {\bf 91}, 150403 (2003).}

\bibitem{qpc}J.\ H.\ Thywissen, R.\ M.\ Westervelt, and M.\ Prentiss,
\emph{Phys.\ Rev.\ Lett.} {\bf 83}, 3762 (1999).

\bibitem{Esteve}J.~Est\`{e}ve, C.~Aussibal, T.~Schumm, C.~Figl,
D.~Mailly, I.~Bouchoule, C.~I.~Westbrook, and A.~Aspect,
\emph{Phys.~Rev.~A} {\bf 70}, 043629 (2004).

\bibitem{Du}S.~Du, M.~B.~Squires, Y.~Imai, L.~Czaia,
R.~A.~Saravanan,V.~Bright, J.~Reichel, T.~W.~H\"ansch, and
D.~Z.~Anderson, \emph{Phys.~Rev.~A} {\bf 70}, 053606 (2004).

\bibitem{ExpReview}See, for intance, W.~Ketterle, D.~S.~Durfee, and
D.~M.~Stamper-Kurn, in {\em Bose-Einstein condensation in atomic
gases, Proceedings of the International School of Physics ``Enrico
Fermi''}, Course CXL, edited by M.~Inguscio, S.~Stringari and
C.~E.~Wieman (IOS Press, Amsterdam, 1999) pp.~67-176 (cond-mat/9904034). 
A more detailed
description of one BEC expriment is given in H.~J.~Lewandowski,
D.~M.~Harber, D.~L.~Whitaker, and E.~A.~Cornell, \emph{J.~Low Temp.~Phys.}
{\bf 132}, 309 (2003).

\bibitem{lifetime}J. E. Bjorkholm, \emph{Phys. Rev. A} {\bf 38}, 1599
(1988).

\bibitem{radiative}{Radiative transfer is unaffected by the vacuum, but not
significant because the absorption frequencies of ground-state alkali
atoms are far above the emission range of a blackbody at room
temperature.}

\bibitem{VacuumNotes} The ion pump is a Varian VacIon Plus 75 StarCell
pump (75~l/s); the sublimation pump uses a Varian 8'' CF Cryopanel; the
turbo pump used is a Pfeiffer Vacuum TMU071P; the gate valve used is a
VAT 10836-UE44-0005 DN63 2-1/2''.

\bibitem{dispensers}C.~Wieman, G.~Flowers, and S.~Gilbert,
\emph{Am.~J.~Phys.}  {\bf 63}, 317 (1995); J.~Fortagh, A.~Grossmann,
T.~W.~H\"{a}nsch, and C.~Zimmermann, \emph{J.\ Appl.\ Phys.} {\bf 84} 6499
(1998).

\bibitem{SAES} Alkali metal dispeners are available from SAES Getters.

\bibitem{K40dispenser}B.~DeMarco, H.~Rohner, and D.~S.~Jin,
\emph{Rev.~Sci.~Instr.} {\bf 70}, 1967 (1999).

\bibitem{LIAD}A. Gozzini, F. Mango, J. H. Xu, G. Alzetta, F. Maccarrone, and R. A. Bernheim,
\emph{Il Nuovo Cimento} {\bf 15}, 709 (1993).

\bibitem{LIADfrequ}M. Meucci, E. Mariotti, P. Bicchi, C. Marinelli, and L. Moi, \emph{Europhys. 
Lett.} {\bf 25}, 639 (1994).

\bibitem{LIAD_MOT}{B.~P.~Anderson and M.~A.~Kasevich, \emph{Phys.~Rev.~A}
{\bf 63}, 023404 (2001).}

\bibitem{potassiumLIAD}{LIAD is also effective for both 
$^{39}$K and $^{40}$K, but we have not studied this efficiency as a function of wavelength.}

\bibitem{DApc} Leaving the dispensers off for several weeks, we
observe a slow decay in the atom number in the MOT, as if a
diminishing amount of $^{87}$Rb is accessible to desorption.  This
behavior has also been observed by Dana Anderson and coworkers at
JILA.

\bibitem{Goldwin}J. Goldwin, S. B. Papp, B. DeMarco, and D. S. Jin,
\emph{Phys. Rev. A} {\bf 65}, 021402 (2002).

\bibitem{dichroicNotes} These dichroic waveplates act as a $\lambda/2$
waveplate for 780~nm and a $\lambda$ waveplate for 767~nm. They were
designed and custom built by OptiSource LLC.

\bibitem{wireNotes} The wire is hollow to allow the flow of
pressurized cooling water through the centre of the wire. It was
manufactured by Wolverine Fabricated Products and insulated by S$\&$W
Wire.

\bibitem{evaporation}W.\ Ketterle and N.\ J.\ van Druten, \emph{Adv.\
At.\ Mol.\ Opt.\ Phys.}  {\bf 37}, 181 (1996).

\bibitem{groth}S.~Groth, P.~Kr\"{u}ger, S.~Wildermuth, R.~Folman,
T.~Fernholz, D.~Mahalu, I.~Bar-Joseph, and J.~Schmiedmayer,
cond-mat/0404141 (2004).

\bibitem{adwinNotes}We use an ADwin Pro sequencer, from 
J\"ager Computergesteuerte Messtechnik GmbH.

\bibitem{dbNotes}Isolation for both inputs and outputs is done by the
NVE IL715-3 quad GMR-based device.  The Texas Instruments
SN64BCT25244NT octal buffer gives a tolerant 50~$\Omega$ output to the
array of BNC connectors.

\bibitem{ADNotes}{The DDS chip we use is the Analog Devices
AD9854, and comes on an evaluation board costing roughly US\$200.  
In our
implementation, an on-board 15.360\,MHz reference
oscillator is multiplied up 19 times to give a 291.84\,MHz sample
frequency.}

\bibitem{camNotes}{MicroPix cameras are used (model M640), with
the window covering the CCD chip removed by the manufacturer.}

\bibitem{DualEvap}K.~E.~Strecker, G.~B.~Partridge, and R.~G.~Hulet,
\emph{Phys.~Rev.~Lett.} {\bf 91}, 080406 (2003).

\bibitem{Feshbach}{E.~Tiesinga, B.~J.~Verhaar, and H.~T.~C.~Stoof,
\emph{Phys.~Rev.~A} {\bf 47}, 4114 (1993); S.\ Inouye, M.~R.~Andrews,
J.~Stenger, H.-J.~Miesner, D.~M.~Stamper-Kurn, and W.~Ketterle,
\emph{Nature} {\bf 392}, 151 (1998); J.~L.~Roberts, N.~R.~Claussen,
J.~P.~Burke, Jr., Chris H.~Greene, E.~A.~Cornell, and C.~E.~Wieman,
\emph{Phys.~Rev.~Lett.} {\bf 81}, 5109 (1998).}

\bibitem{lukinRandom}D.~Wang, M.~D.~Lukin, and E.~Demler,
\emph{Phys.~Rev.~Lett.} {\bf 92}, 076802 (2004).

\end{thebibliography}
\end{document}